\documentclass[11pt]{article}
\usepackage{moriond,epsfig}

\def\NPB{{\em Nucl. Phys.} }
\def\PLB{{\em Phys. Lett.} }
\def\PRL{{\em Phys. Rev. Lett.} }
\def\PRD{{\em Phys. Rev.} }
\def\ZPC{{\em Z. Phys.} }
\def\JPC{{\em J. Phys.} }
\def\APPB{{\em Acta Phys. Polon.} }

\begin{document}
\vspace*{4cm}

\title{Mass Spectrum and the Nature of Neutrinos.}
\author{M. Czakon, J. Gluza, J. Studnik and M. Zra{\l}ek}
\address{Department of Field Theory and Particle Physics, Institute 
of Physics, University of Silesia, Uniwersytecka 4, PL-40-007 Katowice, Poland}
\maketitle\abstracts{Taking as input the best fit solar neutrino anomaly description, MSW LMA, and
the tritium beta decay results we estimate the allowed range of neutrino masses independently of their nature. Adding the
present bound on the effective neutrino mass coming from neutrinoless double beta decay, we
narrow this range for Majorana neutrinos. We complete the discussion by considering future perspectives
on determining the neutrino masses, when the oscillation data will be improved and the next experiments
on $(\beta\beta)_{0\nu}$ and $^3H$ decay give new bounds or obtain concrete life-times or distortions in the 
energy distribution.}

We know much more about neutrino masses than yet a few years ago. 
The observed anomalies in atmospheric,
solar and possibly the LSND neutrino experiments, which we believe are explained
by neutrino oscillations, supplied with the tritium beta decay data give hints
on neutrino masses independently of whether they are Dirac or Majorana particles. Additional constraints on Majorana neutrino
masses come from the fact that no neutrinoless double beta decay has been observed to this day.
In this work we present an up to date analysis and future perspectives of finding the neutrino
mass spectrum without any constraints from theoretical models. We consider only the three neutrino
case (i.e. without considering the LSND anomaly), and the latest best fit solar neutrino problem solution,
the MSW LMA \cite{latest}. The oscillation parameters inferred from atmospheric and solar data are
given in Table~\ref{firsttable}. The four neutrino case and other currently acceptable solutions 
of the solar anomaly are considered elsewhere~\cite{rest}. As there are definitely two scales of $\delta m^2$, $\delta m^2_{atm} \gg
\delta m^2_{sol}$, two possible neutrino mass spectra must be considered. The first, known
as normal mass hierarchy ($A_3$) where $\delta m^2_{sol} = \delta m^2_{21} \ll \delta m^2_{32} \approx \delta
m^2_{atm}$ and the second, inverse mass hierarchy spectrum ($A^{inv}_3$) with $\delta m^2_{sol} = \delta 
m^2_{21} \ll \delta m^2_{atm} \approx -\delta m^2_{31}$. Both schemes are not distinguishable by present
experiments. There is hope that future neutrino factories will do that~\cite{newhope}.

\begin{table}[t]
\caption{The allowed range (95\% of CL) and the best fit values of $\sin^2 2\theta$ and
$\delta m^2$ for the atmospheric neutrino oscillation and the best fit MSW LMA solution of
the solar neutrino problem.\label{firsttable}}
\vspace{0.4cm}
\begin{center}
\begin{tabular}{|c|cc|cc|}
\hline
& \multicolumn{2}{|c|}{Allowed range} & \multicolumn{2}{|c|}{Best fit}  \\
& $\delta m^2 [\; \mbox{\rm eV}^2]$ & $\sin^2 2\theta_{solar}$ 
& $\delta m^2 [\; \mbox{\rm eV}^2]$
& $\sin^2 2\theta_{solar}$ \\
\hline 
Atmospheric neutrinos$^2$ & $(1.5-6)\times 10^{-3}$ & $0.84 - 1$ & $3.5\times 10^{-3}$ & $1$ \\
\hline 
Solar neutrinos (MSW LMA)$^1$& $(1.5-10)\times 10^{-5}$ & $0.3 - 0.92$ & $8\times 10^{-5}$ & $0.66$ \\
\hline
\end{tabular}
\end{center}
\end{table}

\begin{figure}[h]
\begin{center}
\epsfig{figure=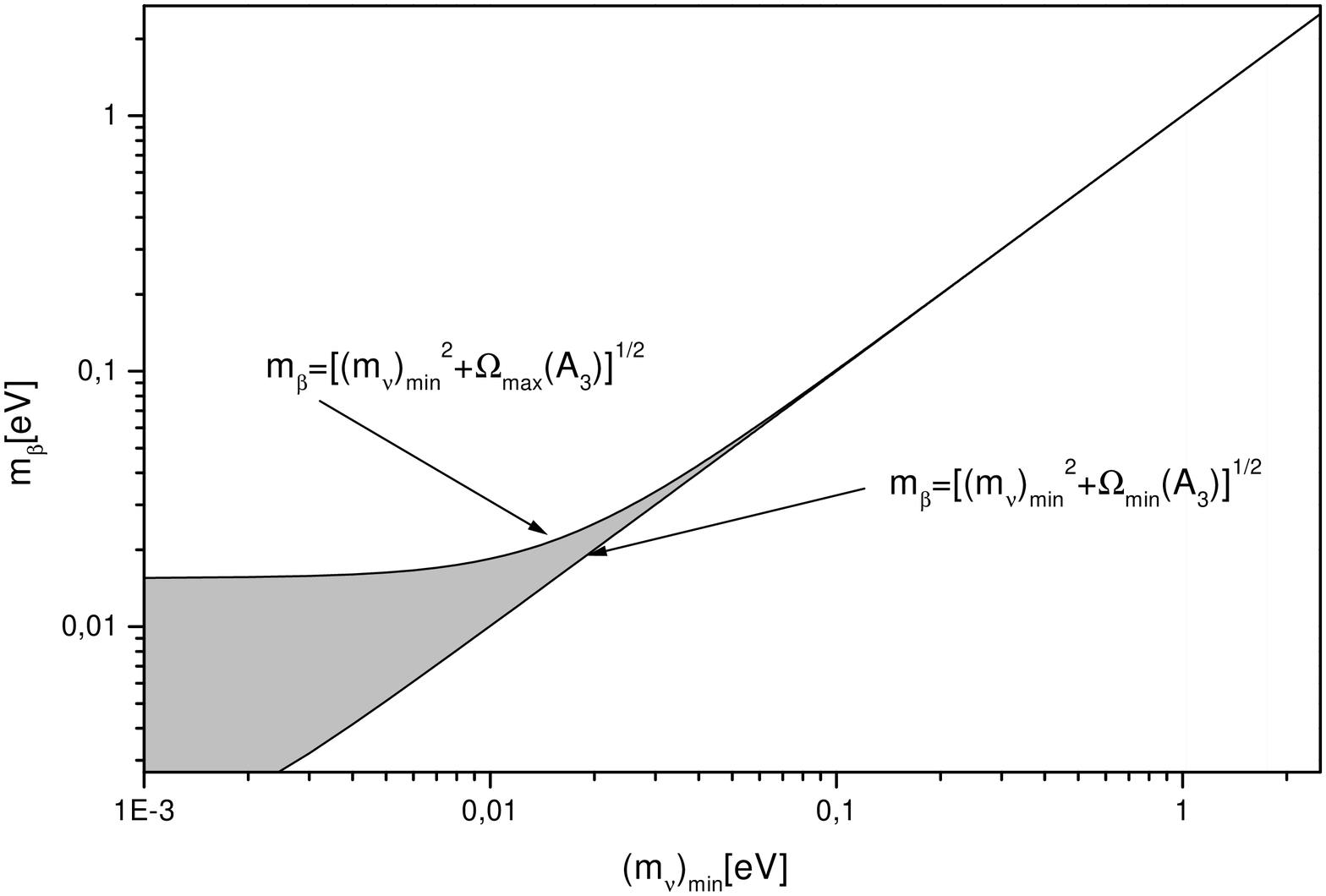, width=7.9cm, height=7cm}
\epsfig{figure=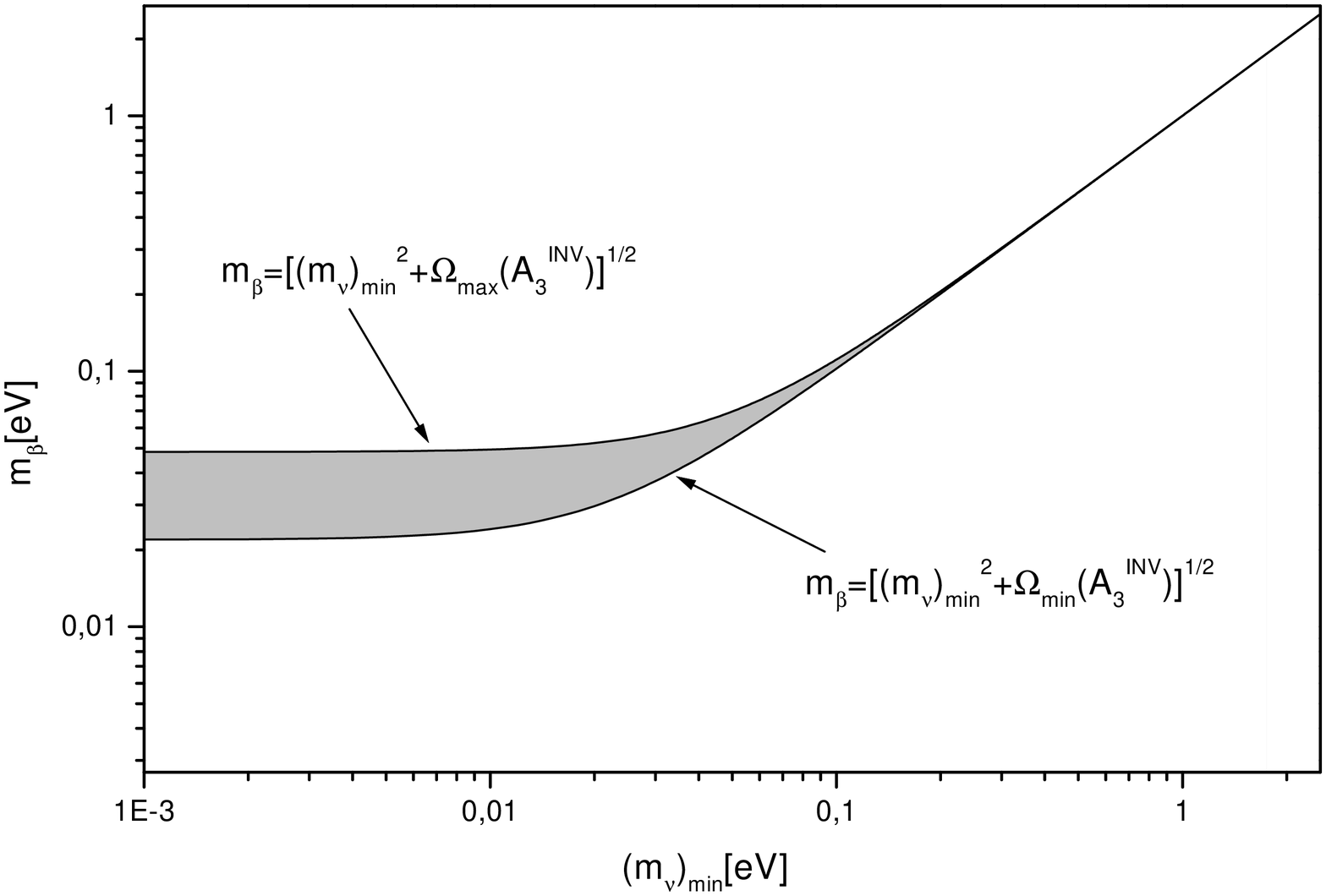, width=7.9cm, height=7cm}
\caption{The allowed range of $m_\beta$ values for a given $(m_\nu)_{min}$ in the $A_3$ scheme (left) and the $A_3^{inv}$ scheme (right).\label{two}}
\end{center}
\end{figure}

Two elements of the first row of the mixing matrix $|U_{e1}|$ and
$|U_{e2}|$ can be expressed by the third element $|U_{e3}|$ and the
$\sin^2{2\theta_{solar}}$
\begin{equation}
|U_{e1}|^2 = (1-|U_{e3}|^2) \frac{1}{2} (1+\sqrt{1-\sin^2{2\theta_{solar}}}),
\end{equation}
and
\begin{equation}
|U_{e2}|^2 = (1-|U_{e3}|^2) \frac{1}{2} (1-\sqrt{1-\sin^2{2\theta_{solar}}}).
\end{equation}
The value of the third element $|U_{e3}|$ is not fixed yet and only
different bounds exist for it. We will take the bound directly
inferred from the CHOOZ and SK experiments \cite{ue3cit}
\begin{equation}
\label{ue3}
|U_{e3}|^2 < 0.04 \;\;\;\; (\mbox{with}\;\; 95\% \;\;\mbox{of CL}).
\end{equation}
Since in both schemes there is
\begin{equation}
(m_\nu)^2_{max} = (m_\nu)^2_{min}+\delta m^2_{solar}+\delta m^2_{atm},
\end{equation}
the oscillation experiments alone give
\begin{equation}
(m_\nu)_{max} \geq \sqrt{\delta m^2_{solar}+\delta m^2_{atm}},
\end{equation}
and
\begin{equation}
|m_i-m_j| \leq \sqrt{\delta m^2_{solar}+\delta m^2_{atm}}.
\end{equation}
Translating the above into numbers (again at $95\%$ CL) \cite{atm} we end up with
\begin{equation}
(m_\nu)_{max} \geq 0.04 \; \mbox{\rm eV}, \;\;\;\; |m_i-m_j| < 0.08\; \mbox{\rm eV}.
\end{equation}
The next important data comes from the tritium beta decay experiments. The
following bound has been lately obtained~\cite{tritium}
\begin{equation}
\label{mbeta}
\left[ \sum^3_{i=1} |U_{ei}|^2m^2_i\right]^{1/2} \equiv m_\beta < \kappa' = 2.2 \; \mbox{\rm eV}
\end{equation}
this obviously leads only to the double inequality
\begin{equation}
(m_\nu)_{min} \leq m_\beta \leq (m_\nu)_{max}.
\end{equation}
Therefore
\begin{equation}
\label{mnumin}
0 \leq (m_\nu)_{min} \leq 2.2\; \mbox{\rm eV}.
\end{equation}
$(m_\nu)_{max}$ remains unfortunately unlimited from above.
Supplying the tritium decay with oscillations we find that~\cite{wazne}
\begin{equation}
\label{omega}
m^2_\beta = (m_\nu)^2_{min}+\Omega_{scheme},
\end{equation}
and
\begin{equation}
(m_\nu)^2_{max} = m^2_\beta+\Lambda_{scheme},
\end{equation}
where $\Omega$ and $\Lambda$ are scheme dependent. For example, in the $A_3$ scheme
\begin{equation}
\Omega(A_3) = (1-|U_{e1}|^2)\delta m^2_{solar} + |U_{e3}|^2 \delta m^2_{atm},
\end{equation}
and
\begin{equation}
\Lambda(A_3) = |U_{e1}|^2 \delta m^2_{solar}+(1-|U_{e3}|^2) \delta m^2_{atm}.
\end{equation}
This provides limits for both $(m_\nu)_{min}$ and $(m_\nu)_{max}$
\begin{equation}
0 \leq  (m_\nu)_{min} \leq \sqrt{(\kappa')^2-\Omega^{min}_{scheme}},
\end{equation}
and
\begin{equation}
\label{big}
\sqrt{\delta m^2_{solar}+\delta m^2_{atm}} \leq (m_\nu)_{max} \leq \sqrt{(\kappa')^2+\Lambda^{max}_{scheme}},
\end{equation}
where this time $\Omega^{min}_{scheme}$ and $\Lambda^{max}_{scheme}$ are the allowed minimal and maximal values.
With the present bound on $m_\beta$ (Eq.~\ref{mbeta}) we recover practically the same range for $(m_\nu)_{min}$ from 
Eq.~\ref{mnumin}, but for $(m_\nu)_{max}$ we obtain from Eq.~\ref{big}
\begin{equation}
0.04 \; \mbox{\rm eV} \leq (m_\nu)_{max} \leq 2.2 \; \mbox{\rm eV}.
\end{equation}
With the help of Eq.~\ref{omega} we plot the range of $m_\beta$ values for a given $(m_\nu)_{min}$ in Fig.~\ref{two} and \ref{three}
for the $A_3$ and $A_3^{inv}$ schemes respectively. We see that the knowledge of $m_\beta$ determines satisfactorily $(m_\nu)_{min}$
for $m_\beta > 0.04\; \mbox{\rm eV} (0.2\; \mbox{\rm eV})$ in the $A_3$($A_3^{inv}$) case. Within this range of $m_\beta$ values the spectrum of neutrino
masses can be determined independently of the neutrino nature (Dirac or Majorana), since none of the above depends on it.
This would be the only possible way to find the masses
if the neutrinos were Dirac particles. In future the value of $m_\beta$ should go down
to $0.5\; \mbox{\rm eV}$~\cite{mbetafuture}. If a value in this range is confirmed, then the spectrum is determined. If not, however, lower
values of $m_\beta$ will require investigation, although this seems to be exteremely difficult.

\begin{figure}
\begin{center}
\epsfig{figure=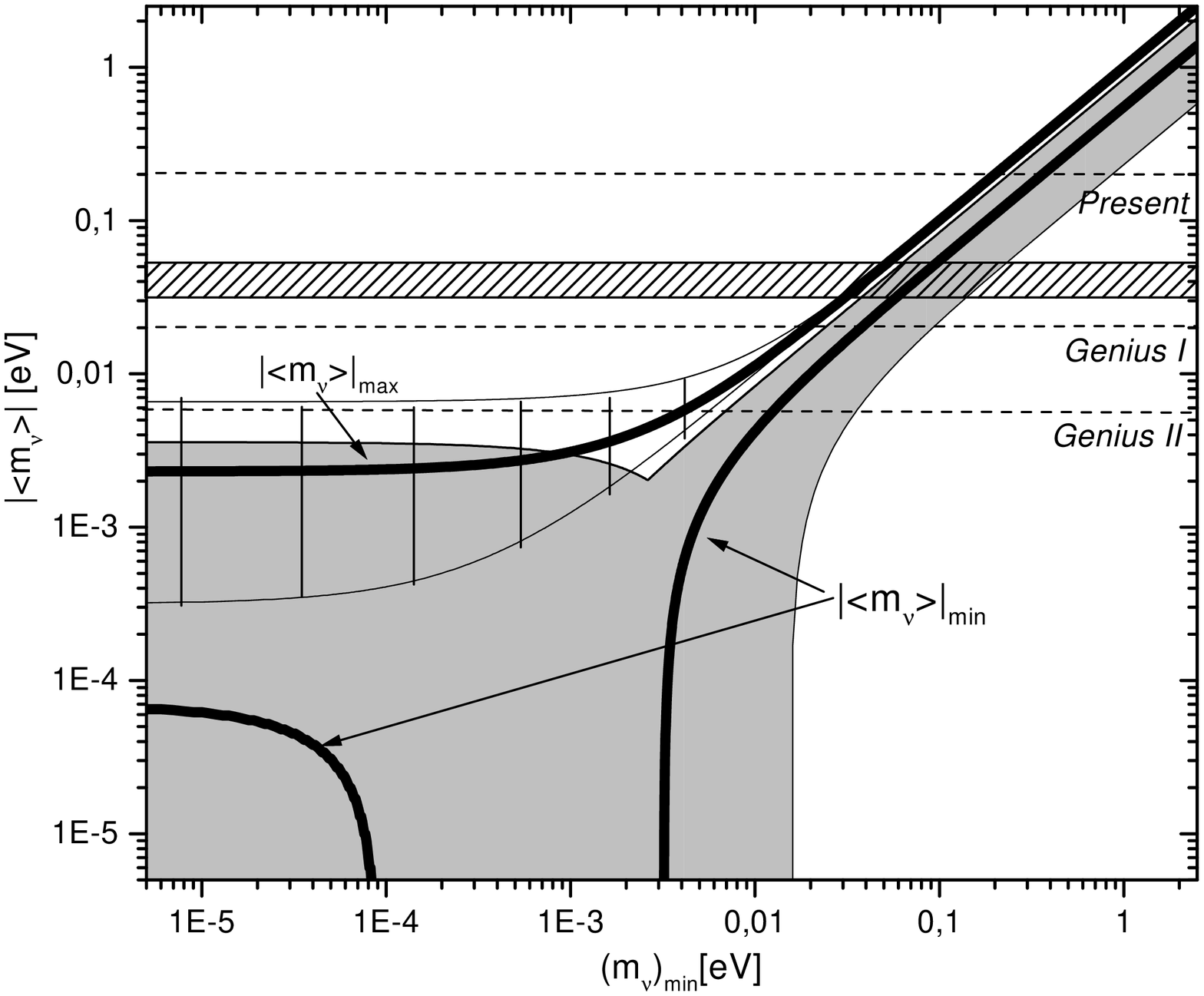, width=12cm}
\caption{Lower $|\langle m_\nu \rangle |_{min}$ and upper $|\langle m_\nu \rangle |_{max}$ limits of the range
of $|\langle m_\nu \rangle |$ as function of $(m_\nu)_{min}$ in the case of the $A_3$ scheme and the best fit
values of the oscillation parameters (solid lines). The
shaded and hashed regions represent the smearing of the limits if the present error bars of the oscillation data
are taken into account. The present and future (GENIUS I and II) bounds on $|\langle m_\nu \rangle |$ are featured. 
The horizontal band  is an example of a GENIUS I positive result with a possible error. 
$|\langle m_\nu \rangle | \in (0.02-0.05)\; \mbox{\rm eV}$. \label{main}}
\end{center}
\end{figure}

For Majorana neutrinos there is one additional constraint, namely the following combination of neutrino masses and mixing
matrix elements can be determined from the neutrinoless double beta decay of nuclei \cite{doi}
\begin{equation}
\langle m_\nu \rangle = \sum_{i=1}^3 U_{ei}^2 m_i.
\end{equation}
The present experiments give only a bound, as no such decay has been observed \cite{presentneutrino}
\begin{equation}
\label{langlemnu}
|\langle m_\nu \rangle | < 0.2 \; \mbox{\rm eV}.
\end{equation}
There are future plans to go down to $|\langle m_\nu \rangle | \simeq 0.02\; \mbox{\rm eV}$ or even  to 
$|\langle m_\nu \rangle | \simeq 0.006\; \mbox{\rm eV}$ \cite{futureneutrino}. Do we have a chance of finding the Majorana mass spectrum
if a value of $|\langle m_\nu \rangle |$ is found within such a small range \cite{pytanie}? This answer as we
will see is not very promising. We shall neglect the difficulties
connected with the determination of $|\langle m_\nu \rangle | $ from the half life time of germanium~\cite{diff}.
As the phases of $U_{ei}$ remain unknown, we are not in position to predict the value of $|\langle m_\nu \rangle | $.
However, the lower $|\langle m_\nu \rangle |_{min}$ and upper $|\langle m_\nu \rangle |_{max}$ ranges as function
of $(m_\nu)_{min}$ can be inferred \cite{nasze}. They are shown in Fig.~\ref{main} for the $A_3$ scheme and
for the MSW LMA solar neutrino problem solution. The shaded and hashed regions give the uncertainties connected with
the allowed ranges of the input parameters ($\sin^2 2\theta_{solar}$, $\delta m^2_{atm}$ (Table~\ref{firsttable}) and
$|U_{e3}|^2$ (Eq.~\ref{ue3}). Future better knowledge of these parameters will reduce the uncertainty regions shown in
Fig.~\ref{main}, but the min-max range caused by the unknown CP phases will remain.

\begin{figure}[h]
\begin{center}
\epsfig{figure=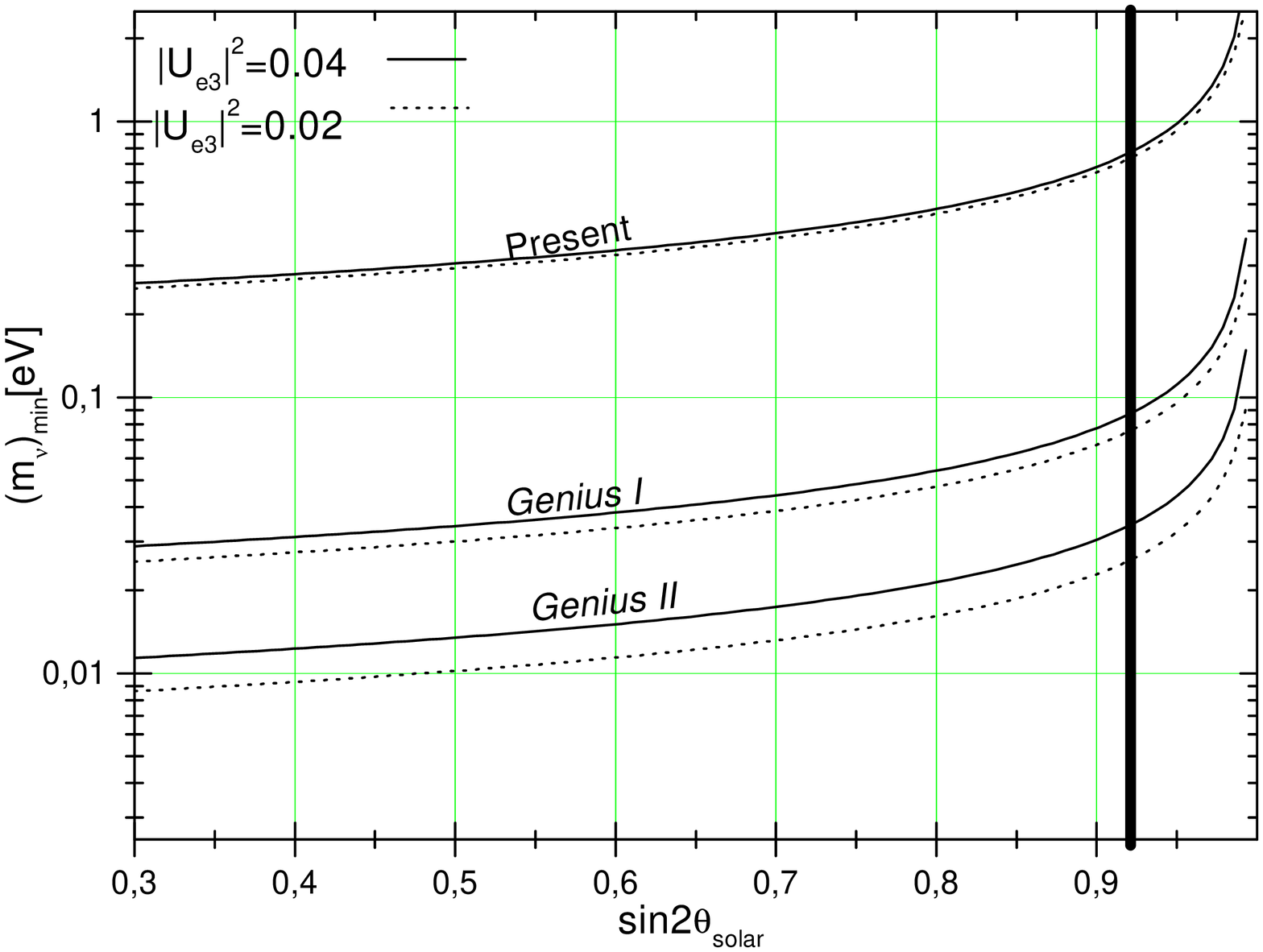, width=12cm}
\caption{The dependence of the bound on $(m_\nu)_{min}$ on $\sin^2 2\theta_{solar}$ for two
different sets of $\delta m^2_{atm}$ and $|U_{e3}|^2$. The different lines represent the values
of $|\langle m_\nu \rangle|$ for different experiments.\label{nextmain}}
\end{center}
\end{figure}

The present experimental bound on $|\langle m_\nu \rangle |$ (Eq.~\ref{langlemnu}) gives the following limit
on the possible $(m_\nu )_{min}$ for Majorana neutrinos
\begin{equation}
\label{limit}
(m_\nu )_{min} < 0.86 \; \mbox{\rm eV}.
\end{equation}
This bound strongly depends on the unknown oscillation parameters, most notably on $\sin^2 2\theta_{solar}$. In
Fig.~\ref{nextmain} we plot this dependence for two different sets of $\delta m^2_{atm}$ and $|U_{e3}|^2$ values.
The limit given in Eq.~\ref{limit} is valid for $\sin^2 2\theta_{solar} = 0.92$, $|U_{e3}|^2 = 0.04$ and
$\delta m^2_{atm} = 6\times 10^{-3}\; \mbox{\rm eV}^2$. If in future, the $(\beta\beta)_{0\nu}$ experiments observe no 
decay, and a new bound is only found, the next better limit that can be derived from Fig.~\ref{nextmain} (with
the present oscillation results), is
\begin{equation}
(m_\nu)_{min} < 0.092 \; \mbox{\rm eV} \;\;\;\; \mbox{GENIUS I},
\end{equation}
and
\begin{equation}
(m_\nu)_{min} < 0.037 \; \mbox{\rm eV} \;\;\;\; \mbox{GENIUS II}.
\end{equation}
In the contrary situation, where a value $|\langle m_\nu \rangle |_{min} \in (0.2 - 0.006)\; \mbox{\rm eV}$ is confirmed, we can try
to predict the Majorana neutrino mass spectrum. The result depends on the value of $|\langle m_\nu \rangle |$ and on
the precision of the oscillation parameters. In Fig.~\ref{main} a possible band of $|\langle m_\nu \rangle |$
values is given with the GENIUS project estimates. The band crosses the region of values allowed by oscillations
giving the possible values of $(m_\nu )_{min}$
\begin{equation}
\label{wazne}
( m_\nu )_{min}^{min (\beta\beta)_{0\nu}} \leq (m_\nu)_{min} \leq (m_\nu)_{min}^{max (\beta\beta)_{0\nu}}.
\end{equation}
With the present day uncertainties on the oscillation parameters, the range of possible values determined by Eq.~\ref{wazne}
is not satisfactorily small. For example, with $|\langle m_\nu \rangle |\simeq 0.05\; \mbox{\rm eV}$
\begin{equation}
\label{haha}
(m_\nu)_{min} \in (0.03-0.6)\; \mbox{\rm eV}.
\end{equation}
For smaller values of $|\langle m_\nu \rangle |$ we can only say that $(m_\nu)_{min} < 0.2\; \mbox{\rm eV}$. A better knowledge of the 
oscillation parameters changes the situation slightly. For example, if the oscillation parameters
are known with negligible error bars for $|\langle m_\nu \rangle |\simeq 0.05\; \mbox{\rm eV}$, then the range Eq.~\ref{haha} changes to
\begin{equation}
(m_\nu)_{min} \in (0.04 - 0.1)\; \mbox{\rm eV}.
\end{equation}
The ignorance of the CP breaking phases in the mixing matrix is fully responsible for this smearing.

\begin{figure}[h]
\begin{center}
\epsfig{figure=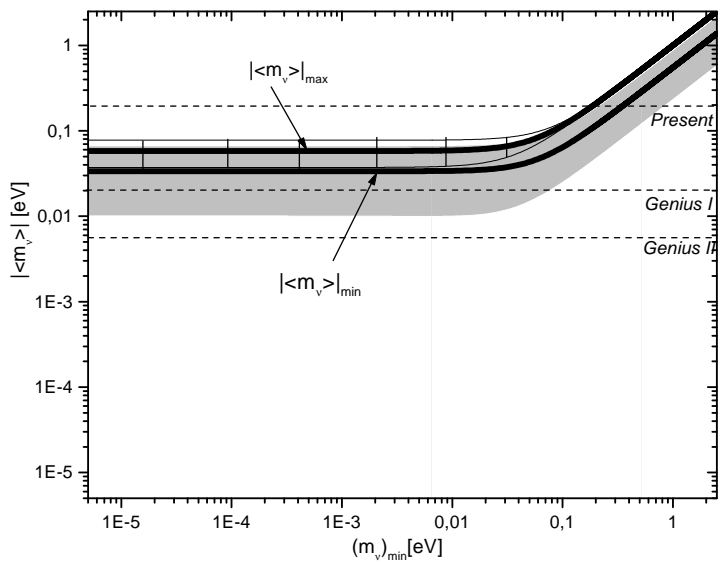, width=12cm}
\caption{Lower $|\langle m_\nu \rangle |_{min}$ and upper $|\langle m_\nu \rangle |_{max}$ limits of the range
of $|\langle m_\nu \rangle |$ as function of $(m_\nu)_{min}$ in the case of the $A_3^{inv}$ scheme and the best fit
values of the oscillation parameters (solid lines). The
shaded and hashed regions represent the smearing of the limits if the present error bars of the oscillation data
are taken into account. The present and future (GENIUS I and II) bounds on $|\langle m_\nu \rangle |$ are featured.
\label{nextnext}}
\end{center}
\end{figure}

The bounds on the effective neutrino mass $|\langle m_\nu \rangle |$ in the inverse hierarchy mass scheme $A_3^{inv}$ and
the MSW LMA solution of the solar neutrino problem  are depicted in Fig.~\ref{nextnext}. We see that the present bound
on $|\langle m_\nu \rangle |$ (Eq.~\ref{langlemnu}), gives a similar limit on the possible range of 
$(m_\nu)_{min}$ of Majorana neutrino masses
\begin{equation}
(m_\nu)_{min} < 0.86 \; \mbox{\rm eV}.
\end{equation}
The first stage of GENIUS can yield
\begin{equation}
(m_\nu)_{min} < 0.077 \; \mbox{\rm eV},
\end{equation}
while the second would exclude the $A_3^{inv}$ scheme.

In conclusion, the present data allow for the following statements
\begin{itemize}
\item we are not able to distinguish between Dirac and Majorana neutrinos.
\item the allowed range of masses for Dirac neutrinos is wider than for Majorana, but
the latter depends strongly on the oscillation parameters.
\item the oscillation and tritium beta decay experiments are able to determine the spectrum
of neutrino masses for values of $m_\beta$ which differ in the $A_3$ ($m_\beta \geq 0.04\; \mbox{\rm eV}$) and
the $A_3^{inv}$ ($m_\beta \geq 0.2\; \mbox{\rm eV}$) schemes.
\item the oscillation and $(\beta\beta)_{0\nu}$ experiments are able to find the range of possible
$(m_\nu)_{min}$ values. However, this range is not small even with
oscillation parameters of negligible error bars. 
\end{itemize}

\section*{Acknowledgments}
One of us (MZ) would like to thank all the organizers and especially Prof. Tran Tanh Van for invitation
and a very good atmosphere at the perfectly prepared conference.
This work was supported by the Polish Committee for Scientific Research under 
Grant No.~2P03B05418 and 2P03B04919.


\section*{References}
\begin{thebibliography}{40}
\bibitem{latest} N.Hata, P.Langacker, \PRD{\bf D56},  6107 (1997);
J.N.Bahcall, P.Krastev, A.Yu.Smirnov, \PRD{\bf D58},  096016 (1998); 
V. Banger, K. Whisnant, \PRD {\bf D59},  093007 (1999); 
M.C. Gonzalez-Garcia, P.C. de Holanda, C.Pena-Garay, J.W.J. Valle,
hep-ph/9906469; 
A. de Gouvea, A. Friedland, A.Murayama, hep-ph/0002064; 
M.G. Gonzalez-Garcia, C. Pena-Garay, hep-ph/0002186;
G.~L.~Fogli, E.~Lisi, D.~Montanino and A.~Palazzo,
hep-ph/9912231;
M.C. Gonzalez-Garcia, C. Pena-Garay, hep-ph/00099041.
\bibitem{atm} Y. Fukuda et al., \PLB{\bf B433},  9 (1998);
\PLB{\bf B436},  33 (1998);\PRL{\bf 81},  1562 (1998);
\PRL{\bf 82},  2644 (1999);
W. A. Mann, hep-ex/9912007; A. De Rujula, M. B. Gavela, P. Hernandez, hep-ph/0001124;
N. Fornengo, M. C. Gonzalez-Garcia, J. W. F. Valle, hep-ph/0002147;
S. Fukuda et al. (Superkamiokande Coll.), hep-ex/00090001;
G.L. Fogli, E. Lisi, A. Marrone, D. Nontanino, hep-ph/0009269.
\bibitem{rest} M. Czakon, J. Studnik, M. Zra{\l}ek, hep-ph/0006339;
M. Czakon, J. Gluza, M. Zra{\l}ek, hep-ph/0003161.
\bibitem{newhope} SuperKamiokande homepage, http://www-sk.icrr.u-tokyo.ac.jp/doc/sk;
SNO homepage, http://snodaq.phy.queensu.ca/SNO/sno.html;
BOREXINO homepage, http://almime.mi.infn.it;
HERON homepage, http://www.physics.brown.edu/research/heron;
HELLAZ homepage, http://sg1.fsu.edu/hellaz;
K. Nishikawa, \NPB (Proc. Supp.) {\bf 77},  198 (1999);
B. C. Barish, \NPB (Proc. Supp.) {\bf 70},  227 (1999);
A. Cervera et al., hep-ph/0002108;
V. Barger, S. Geer, R. Raja, K. Whisnant, hep-ph/0007181;
S. Geer, hep-ph/0008155;
R. Burton, hep-ph/008222.
\bibitem{ue3cit} G.L. Fogli, E. Lisi, A. Marrone, G. Scioscia, \PRD {\bf D59},  033001 (1999);
CHOOZ coll., \PLB {\bf B466},  415. (1999)
\bibitem{tritium}  C. Weinheimer et al., \PLB {\bf B460}
,  219 (1999);
V. M. Lobashev et al., \PLB {\bf B460}
,  227 (1999);
Mainz Collaboration, Neutrino 2000, Canada.
\bibitem{wazne} compare V.~Barger and K.~Whisnant,
\PLB  {\bf B456},  54., (1999) 
S.~Goswami, D.~Majumdar and A.~Raychaudhuri,
hep-ph/9909453.
\bibitem{mbetafuture} K. Zuber, hep-ph/9911362.
\bibitem{doi} see e.g. M. Doi, T. Kotani, E. Takasugi, Prog. Theor. Phys.
(supplement) {\bf 83},  1. (1985)
\bibitem{presentneutrino} L. Baudis et al., \PRL {\bf 83}, 
41. (1999)
\bibitem{futureneutrino} H. V. Klapdor-Kleigrothaus, hep-ex/9907040;
L. Baudis et al., GENIUS (Collaboration), hep-ph/9910205.
\bibitem{pytanie} S.T. Petcov, A. Y. Smirnov, \PLB {\bf B322},  109 (1994); S.M. Bilenky, A. Bottino, 
C. Giunti, C. Kim, \PRD {\bf D54},  1881 (1996); S.M. Bilenky, C. Giunti, C. Kim, S. Petcov, \PRD {\bf D54} 
,  4444 (1996); J. Hellmig, H.V. Klapdor-Kleingrothaus, \ZPC {\bf A359},  351 (1997); H. V. Klapdor-Kleingrothaus, 
J. Hellmig, M. Hirsch, \JPC {\bf G24},  483 (1998); H. Minataka, O. Yasuda, \PRD {\bf D56},  1692, (1997) 
\NPB {\bf  B523},  597 (1998); S.M. Bilenky, C. Giunti, C. W. Kim, M. Monteno, \PRD {\bf D54},  6981, (1998) 
hep-ph/9904328; F. Vissani, hep-ph/9708482, hep-ph/9904349, hep-ph/9906525; T. Fukuyama, K. Matsuda, 
H. Nishiura, hep-ph/9708397; {\em Mod. Phys. Lett.} {\bf A13},  2279 (1998); S.M. Bilenky, C. Giunti, W. Grimus, hep-ph/9809368; 
S. Bilenky, C. Giunti, hep-ph/9904328; S. Bilenky, C. Giunti, W. Grimus, B. Kayser, S.T. Petcov, hep-ph/9907234,
\PLB {\bf B465},  193 (1999);
H. Georgi, S.L. Glashow, hep-ph/9808293; V. Barger, K. Whisnant, \PLB B{\bf 456},  194 (1999); J. Ellis, S. Lola, 
hep-ph/9904279, \PLB {\bf B458},  310 (1999); G.C. Branco, M. N. Rebelo, J.I. Silva-Marcos, \PRL {\bf 82},  683 (1999);
C. Giunti, \PRD {\bf D61},  036002 (2000); R. Adhikari, G. Rajasekaran, hep-ph/9812361; \PRD{\bf D61},  031301 (2000); 
K. Matsuda, N. Takeda, T. Fukuyama, H. Nishiura, hep-ph/0003055; 
M.~Czakon, M.~Zralek and J.~Gluza,
\APPB  {\bf B30},  3121 (1999)
M.~Czakon, J.~Gluza and M.~Zralek,
\PLB  {\bf B465},  211. (1999)
\bibitem{diff} H.V. Klapdor-Kleingrothaus, H. Paes, A. Yu. Smirnov, hep-ph/0003219;
\bibitem{nasze} 
M.~Czakon, J.~Studnik, M.~Zralek and J.~Gluza,
\APPB  {\bf B31},  1365 (2000);
\end{thebibliography}
\end{document}